\documentclass{PoS}
\usepackage{url}

\newcommand{\eqn}[1]{(\ref{#1})}

\newcommand{\ovl}[1]{\overline{#1}}

\newcommand{\vev}[1]{\left\langle #1 \right\rangle}

\title{Study of energy-momentum tensor correlation function in $N_{f}=2+1$ full QCD for QGP viscosities}

\ShortTitle{Study of energy-momentum tensor correlation function}

\author{\speaker{Yusuke Taniguchi}
\\
Center for Computational Sciences, University of Tsukuba,
Tsukuba, Ibaraki 305-8571, Japan
\\
        E-mail: \email{tanigchi@het.ph.tsukuba.ac.jp}}

\author{Atsushi Baba, Asobu Suzuki
\\
Graduate School of Pure and Applied Sciences, University of Tsukuba,
Tsukuba, Ibaraki, Japan
        }

\author{Shinji Ejiri
\\
Department of Physics, Niigata University, Niigata 950-2181, Japan
        }

\author{Kazuyuki Kanaya
\\
Tomonaga Center for the History of the Universe, University of Tsukuba, Tsukuba, Ibaraki 305-8571, Japan
        }

\author{Masakiyo Kitazawa
\\
Department of Physics, Osaka University, Osaka 560-0043, Japan
\\
J-PARC Branch, KEK Theory Center, Institute of Particle and Nuclear Studies,
KEK, 203-1, Shirakata, Tokai, Ibaraki, 319-1106, Japan
        }

\author{Takanori Shimojo, Hiroshi Suzuki
\\
Department of Physics, Kyushu University, 744 Motooka, Nishi-ku, Fukuoka 819-0395, Japan
        }

\author{Takashi Umeda
\\
Graduate School of Education, Hiroshima University,
Higashihiroshima, Hiroshima 739-8524, Japan
        }

\abstract{
We study correlation functions of the energy-momentum tensor (EMT) in $(2+1)$-flavor full QCD to evaluate QGP viscosities.
We adopt nonperturbatively improved Wilson fermion and Iwasaki gauge action.
Our degenerate $u$, $d$ quark mass is rather heavy with $m_{\pi}/m_{\rho}\simeq0.63$,
while the $s$ quark mass is set to approximately its physical value.
Performing simulations on lattices with $N_t=16$ to 6 at a fine lattice spacing of $a=0.07$ fm, the temperature range of $T\simeq174$--$464$ MeV is covered using the fixed-scale approach.
We attempt to compute viscosities by three steps: 
(1) calculate two point correlation functions of non-perturbatively renormalized EMT applying the gradient flow method,
(2) derive the spectral function from correlation function, and 
(3) extract viscosities from the spectral function applying the Kubo formula.
We report on the status of the project and present preliminary results for the shear viscosity in the high temperature phase.
\\
Preprint numbers: UTCCS-P-120, UTHEP-730, KYUSHU-HET-191, J-PARC-TH-0153
}

\FullConference{The 36th Annual International Symposium on Lattice Field Theory - LATTICE2018\\
		22-28 July, 2018\\
		Michigan State University, East Lansing, Michigan, USA.}

\begin{document}

\section{Introduction}

Ever since the heavy ion collision experiments went over the QCD phase transition temperature, the physical property of the quark gluon plasma (QGP) has been one of the most fascinating topics in high energy physics.
The discovery of the large elliptic flow indicates that a kind of collective motion exists in QGP, and
 and the success of the hydrodynamic models suggests that the QGP near the critical temperature is in
 a strongly coupled liquid state rather than a gas state
(for a review see \cite{Akiba:2015jwa} and references there in).
Furthermore, the comparison between hydrodynamic models and experimental data indicates that
the shear viscosity of the QGP is quite small with $\eta/s\sim0.2$, suggesting that the QGP liquid found is the most perfect fluid we have ever seen
 (for a review see Ref.~\cite{Gale:2013da}).

Inspired by the progress of experiment, there have been challenges in the lattice QCD community to calculate viscosities of QGP \cite{Nakamura:2004sy,Meyer:2007ic,Meyer:2009jp,Pasztor:2018yae,Astrakhantsev:2018oue}.
Viscosities can be extracted from the correlation function of the energy-momentum tensor (EMT) applying Kubo's linear response relations.
There are two major difficulties in this procedure to be carried out on the lattice.
Firstly, the lack of translational symmetry on the lattice makes a definition of EMT not trivial.
Secondly, the Kubo's response function is defined in the real time formalism and thus not directly calculable in the Euclidean space time.
We need to use the spectral function to relate the Matsubara Green's function in the imaginary time formalism to the Kubo's response function.
However, a derivation of the continuous spectral function is an ill-posed problem on a finite lattice.
Because of the these difficulties, studies of viscosity on the lattice have been limited to the pure SU(3) theory without quarks.

We solve the first difficulty by adopting the gradient flow method
\cite{Narayanan:2006rf,Luscher:2010iy,Luscher:2013cpa}
as a non-perturbative renormalization scheme for the EMT \cite{Suzuki:2013gza,Makino:2014taa}.
In Ref.~\cite{Taniguchi:2016ofw}, we have computed the one-point function of EMT in $N_{f}=2+1$ full QCD with the gradient flow method, 
and found that the results for the diagonal elements of EMT are well consistent with the equation of state evaluated previously with the conventional $T$-integration method.
We now compute two-point functions of diagonal as well as off-diagonal elements of EMT.
For the second difficulty, we try a couple of ans\"atze for the spectral function to fit the correlation functions in this study.
By virtue of the gradient flow the correlation function is evaluated with a good precision and the fit works well.

\section{Viscosities}

The shear viscosity $\eta$ is given by a proportional coefficient in the stress tensor (spatial components of EMT) against a gradient of the velocity field $u_{i}$ for sufficiently long wavelength and slow mode
\begin{eqnarray}
&&
\langle T_{ij}\rangle_{\beta}=-\eta\partial_{i}u_{j},\quad
(i\neq j),
\end{eqnarray}
where
\begin{eqnarray}
&&
\vev{T_{ij}}_{\beta}=\frac{{\rm Tr}\left(\exp\left(-\beta{H}\right)T_{ij}\right)}
{{\rm Tr}\exp\left(-\beta{H}\right)}
\end{eqnarray}
with $\beta$ the inverse temperature and $H$ the Hamiltonian.
The bulk viscosity $\zeta$ is given in a similar manner, $\langle T_{ii}\rangle_{\beta}=-\zeta\partial_{i}u_{i}$ 
where $i$ is not summed over.

In thermal equilibrium, however, the medium is isotropic and we have $\langle T_{ij}\rangle_{\beta}=0$ for $i\neq j$.
This is consistent with a fact that hydrodynamics is a non-equilibrium phenomenon.
We thus introduce velocity field $u_{l}(\vec{x})$ to study a  relaxation process due to local fluctuation  of the momentum density $T^{0l}$ interns of time dependent stress tensor:
\begin{eqnarray}
&&
\langle{T}_{ij}(t,\vec{x})\rangle_{u}
=\frac{1}{Z_{u}}{\rm Tr}\left(
\exp\left(-\beta{H}+\beta\int d^{3}x\,{T^{0l}}(\vec{x})u_{l}(\vec{x})\right){T}_{ij}(t,\vec{x})\right),
\\&&
Z_{u}={{\rm Tr}\exp\left(-\beta{H}+\beta\int d^{3}x\,{T^{0l}}(\vec{x})u_{l}(\vec{x})\right)},
\end{eqnarray}
where $(t,\vec{x})$ is a coordinate of the Minkowski space-time with $\eta_{\mu\nu}=(+---)$.
Using a relation
\begin{eqnarray*}
&&
\exp\left(-\beta{H}+\beta\int d^{3}x\,{T^{0l}}(\vec{x})\,u_{l}(\vec{x})\right)
=\exp\left(-\beta{H}\right)
T_{\tau}\exp\left(\int_{0}^{\beta}d\tau \int d^{3}x\,{T^{0l}}(-i\tau,\vec{x})\,u_{l}(\vec{x})\right)
\end{eqnarray*}
with
${T^{0l}}(-i\tau,\vec{x})=e^{\tau{H}}{T^{0l}}(\vec{x})e^{-\tau{H}}$
and $T_\tau$ for the $\tau$-ordered product operation, 
we expand the expectation value to the leading order in $u_{l}$,
\begin{eqnarray}
\langle{T}_{ij}(t,\vec{x})\rangle_{u}&=&
\int_0^{\beta}d\tau \int d^{3}x'
\vev{
\Delta{T}_{ij}(t-i\tau,\vec{x})\,
\Delta{T^{0l}}(0,\vec{x}')
}_\beta
u_{l}(\vec{x}'),
\end{eqnarray}
where
$\Delta T_{ij}=T_{ij}-\vev{T_{ij}}_{\beta}$.

Making use of the conservation of EMT,
the translation invariance and the time reversal symmetry in real time, we have
\begin{eqnarray}
\langle{T}_{ij}(t,\vec{x})\rangle_{u}&=&
-\int_{0}^{t}ds \int d^{3}x'\int_{0}^{\beta}d\tau
\langle\Delta{T}_{ij}(s-i\tau,\vec{x})\,
\Delta{T}^{kl}(0,\vec{x}')\rangle_{\beta}\,\partial_{k}u_{l}(\vec{x}').
\end{eqnarray}
Assuming a uniform gradient for the velocity field $u_{l}(\vec{x})$ we extract the viscosity.
Further applying the translational invariance in space-time we finally get linear response relations for the viscosities
\begin{eqnarray}
&&
\eta=
\int_{0}^{\infty}dt \int d^{3}x\int_{0}^{\beta}d\tau\,
\langle\Delta{T}_{ij}(t-i\tau,\vec{x})\,
\Delta{T}_{ij}(0,\vec{0})\rangle_{\beta},
\label{eq:eta}
\\&&
\zeta=
\int_{0}^{\infty}dt \int d^{3}x\int_{0}^{\beta}d\tau
\langle\Delta{T}_{ii}(t-i\tau,\vec{x})\,
\Delta{T}_{ii}(0,\vec{0})\rangle_{\beta},
\label{eq:zeta}
\end{eqnarray}
where sum is not taken over spatial indices.

The Kubo's canonical correlation in the right hand sides of \eqn{eq:eta} and \eqn{eq:zeta} is defined in the real time formalism and is not simply available on the lattice with Euclidean time.
We use the spectral function to relate Kubo's response function to the Matsubara Green function in Euclidean time.
The spectral function is defined as usual
\begin{eqnarray}
&&
\rho_{ij;kl}(k)\equiv\int d^{4}x\,e^{ikx}\left\langle\left[{T_{ij}}(x),{T_{kl}}(0)\right]\right\rangle_\beta.
\end{eqnarray}
It is easy to show that both the Kubo and Matsubara Green functions are related with the spectral function as follows
\begin{eqnarray}
&&
\int d^{3}x\,\langle{T_{ij}}(-i\tau,\vec{x}){T_{kl}}(0,\vec{0})\rangle_{\beta}
=\int_{0}^{\infty}\frac{dk_{0}}{2\pi}\frac{\cosh k_{0}\left(\tau-\frac{\beta}{2}\right)}{\sinh k_{0}\frac{\beta}{2}}\rho_{ij;kl}(k_{0},\vec{0}),
\label{euclidean-spectral-function}
\\&&
\int_{0}^{\beta}d\tau \int_{-\infty}^\infty dt\,e^{ik_{0}t}
\int d^{3}x\,e^{-i\vec{k}\vec{x}}
\langle{T_{ij}}(t-i\tau,\vec{x})\,{T_{kl}}(0,\vec{0})\rangle_{\beta}
=\frac{\rho_{ij;kl}(k)}{k_{0}}.
\end{eqnarray}
When the spectral function is derived from the Euclidean correlation functions according to \eqn{euclidean-spectral-function}, the viscosities are given by 
\begin{eqnarray}
\eta=\lim_{k_0\to0}\frac{\rho_{ij;ij}(k_0,,\vec{0})}{2k_0},
\quad\quad
\zeta=\lim_{k_0\to0}\frac{\rho_{ii;ii}(k_0,,\vec{0})}{2k_0}.
\end{eqnarray}

\section{Lattice calculation}
\label{sec-gradient-flow}

We compute nonperturbatively renormalized EMT by the gradient flow method of Refs.~\cite{Suzuki:2013gza,Makino:2014taa}.
We adopt the flow equations given in Refs.~\cite{Luscher:2010iy,Luscher:2013cpa}
for the gauge and quark fields
\begin{eqnarray}
&&
   \partial_tB_\mu(t,x)=D_\nu G_{\nu\mu}(t,x),
   \quad
   B_\mu(t=0,x)=A_\mu(x),
\label{eq:(2.1)}
\\&&
\partial_t\chi_f(t,x)=D_\mu D_\mu\chi_f(t,x),
\quad
   \chi_f(t=0,x)=\psi_f(x),
\label{eq:(2.4)}
\\&&
\partial_t\ovl{\chi}_f(t,x)
   =\ovl{\chi}_f(t,x)\overleftarrow{D}_\mu\overleftarrow{D}_\mu,
   \quad
   \ovl{\chi}_f(t=0,x)=\ovl{\psi}_f(x),
\label{eq:(2.5)}
\end{eqnarray}
where the field strength $G_{\mu\nu}$ and the covariant derivative $D_\mu$ are given in terms of the flowed gauge field $B_\mu$, 
and $f=u$, $d$, $s$, denotes the flavor index.
The properly normalized EMT which satisfies the Ward-Takahashi identity in the continuum limit is given by two steps.
First, we make a flowed tensor operator $T_{\mu\nu}(t,x)$ using flowed fields \eqn{eq:(2.1)}-\eqn{eq:(2.5)}
by a linear combination of five operators $\tilde{O}_{i\mu\nu}(t,x)$
multiplied with matching coefficients $c_{i}(t)$ defined in Refs.~\cite{Suzuki:2013gza,Makino:2014taa}.
Then, we take the limit $T_{\mu\nu}(x)=\lim_{t\to0}T_{\mu\nu}(t,x)$ 
to resolve mixing with irrelevant dimension six operators.
%
To carry out the $t\to0$ extrapolation on finite lattices avoiding $O(a^2/t)$ singularities at small $t$, we adopt the strategy of Ref.~\cite{Taniguchi:2016ofw}, 
\textit{i.e.}, we first identify a region of $t$ (linear window) in which a linear behavior is visible, and then extrapolate the data by a linear fit in the linear window.
%
Some numerical techniques to calculate the correlation function 
$\langle{T_{ij}}(t,{x}){T_{kl}}(t,0)\rangle_{\beta}$ at flow time $t$ are explained in Ref.~\cite{Taniguchi:2017ibr}.


The spectral function $\rho_{ij;kl}(k)$ is a continuous function in four momentum $k$.
It is an ill-posed problem to derive it from finite-number data of lattice correlation function obtained on a lattice with finite volume.
In this study, we try to get the spectral function by a fit to the correlation function using two types of fit ans\"atze.
One is the Breit-Wigner form for the spectral function \cite{Nakamura:2004sy}
\begin{eqnarray}
&&
\frac{\rho_{\rm BW}(k_{0})}{k_{0}}=\frac{F}{1+b^{2}(k_{0}-\omega_{0})^{2}}
+\frac{F}{1+b^{2}(k_{0}+\omega_{0})^{2}},
\label{BW-model}
\end{eqnarray}
where $F$, $b$ and $\omega_0$ are fit parameters.
The other is the hard thermal loop ansatz which uses a combination of the hydrodynamical model and perturbative prediction
\begin{eqnarray}
&&
\frac{\rho_{\rm HTL}(k_{0})}{k_{0}}=\frac{a}{1+b^{2}k_{0}^{2}}
+\theta(k_{0}-\omega_{0})\frac{Ak_{0}^{3}}{\tanh\frac{k_{0}}{4T}},
\label{HTL-model}
\end{eqnarray}
where $a$, $b$, $\omega_0$ and $A$ are fit parameters.

\section{Numerical results}

Measurements of EMT with the gradient flow method are performed on the $N_f=2+1$ gauge
configurations generated for Ref.~\cite{Umeda:2012er}.
The zero-temperature gauge configurations were generated for Ref.~\cite{Ishikawa:2007nn}.
The nonperturbatively $O(a)$-improved Wilson quark action and
the renormalization-group improved Iwasaki gauge action are adopted.
The bare coupling constant is set to $\beta=2.05$, which corresponds
to~$a=0.0701(29)\,\mathrm{fm}$ ($1/a\simeq2.79\,\mathrm{GeV}$).
The hopping parameters are set to $\kappa_u=\kappa_d\equiv\kappa_{ud}=0.1356$
and $\kappa_s=0.1351$, which correspond to heavy $u$ and $d$ quarks,
$m_\pi/m_\rho\simeq0.63$, and almost physical $s$ quark, $m_{\eta_{ss}}/m_\phi\simeq0.74$.
We adopt the fixed-scale approach~\cite{Umeda:2008bd}  in which the temperature $T=1/(aN_t)$ is varied by changing the temporal lattice size $N_t$ with a fixed lattice spacing $a$.
The temperature varies $174\le T\le464$ MeV with temporal length $16\ge N_{t}\ge6$.
See Ref.~\cite{Taniguchi:2016ofw} for a detailed explanation of numerical parameters.

\begin{figure}[tb]
 \centering
  \includegraphics[width=4.9cm,trim=-20mm -10mm 0mm 0mm,clip]{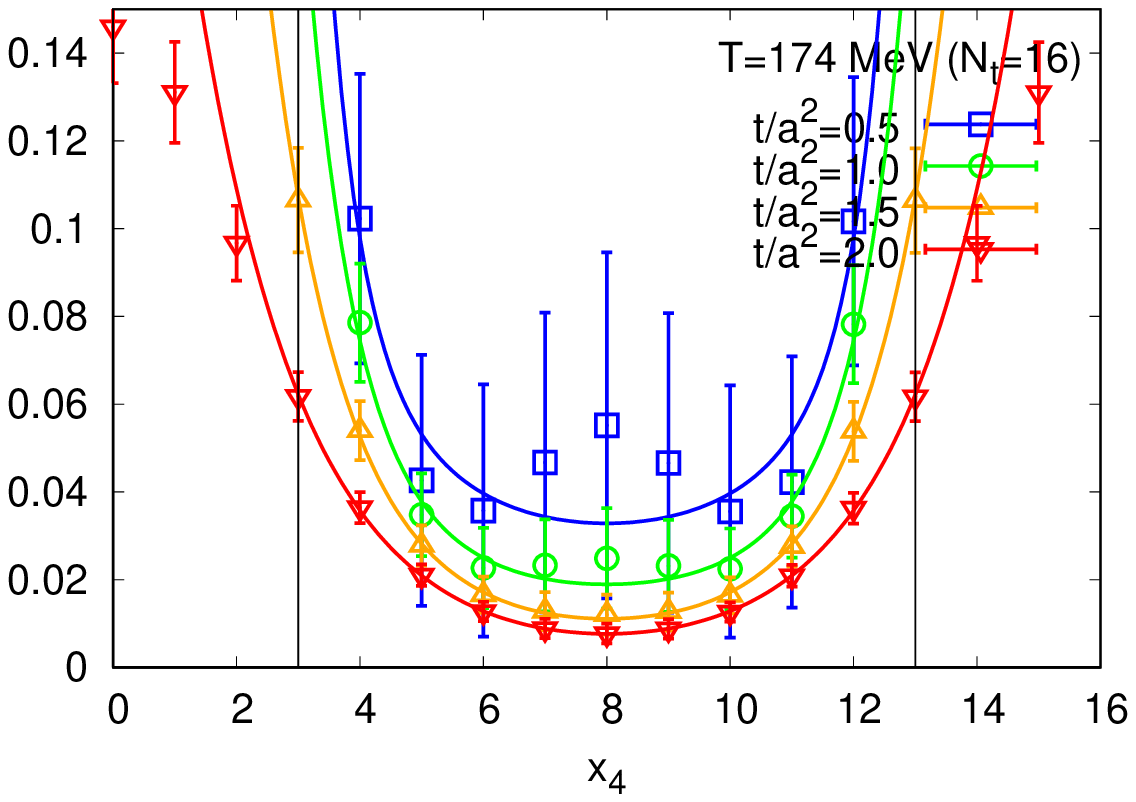}
  \includegraphics[width=4.9cm,trim=-20mm -10mm 0mm 0mm,clip]{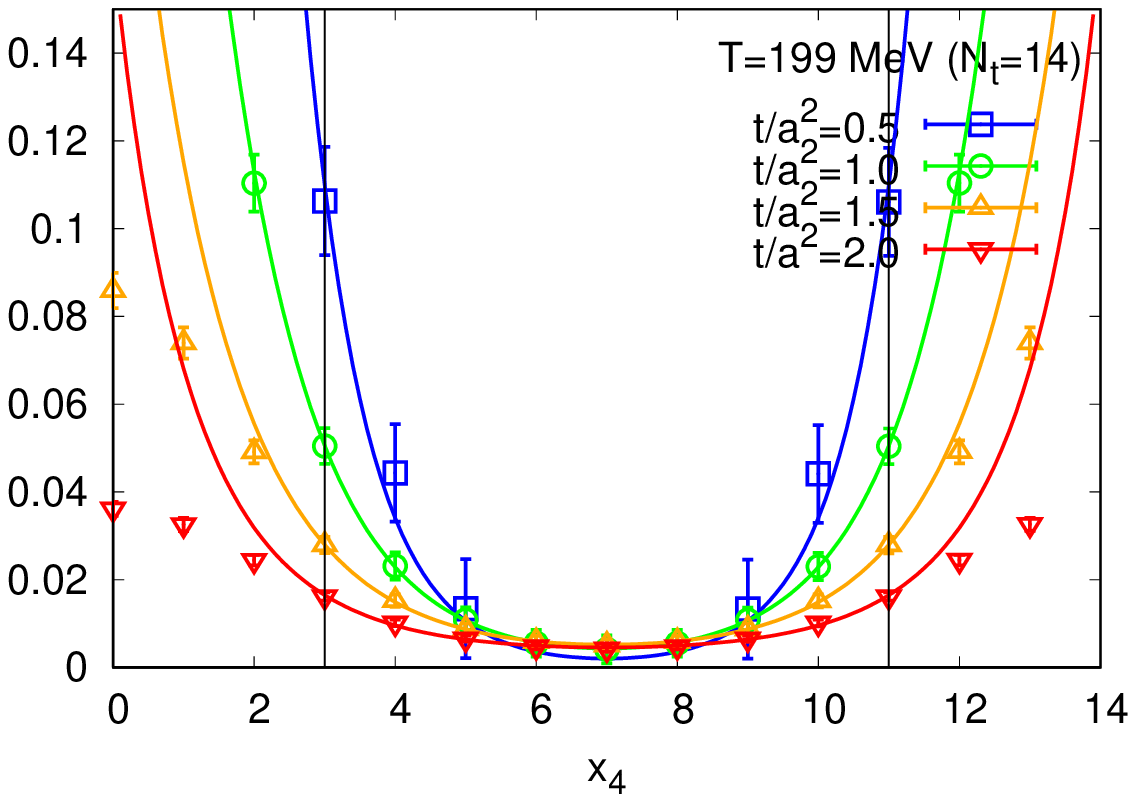}
  \includegraphics[width=4.9cm,trim=-20mm -10mm 0mm 0mm,clip]{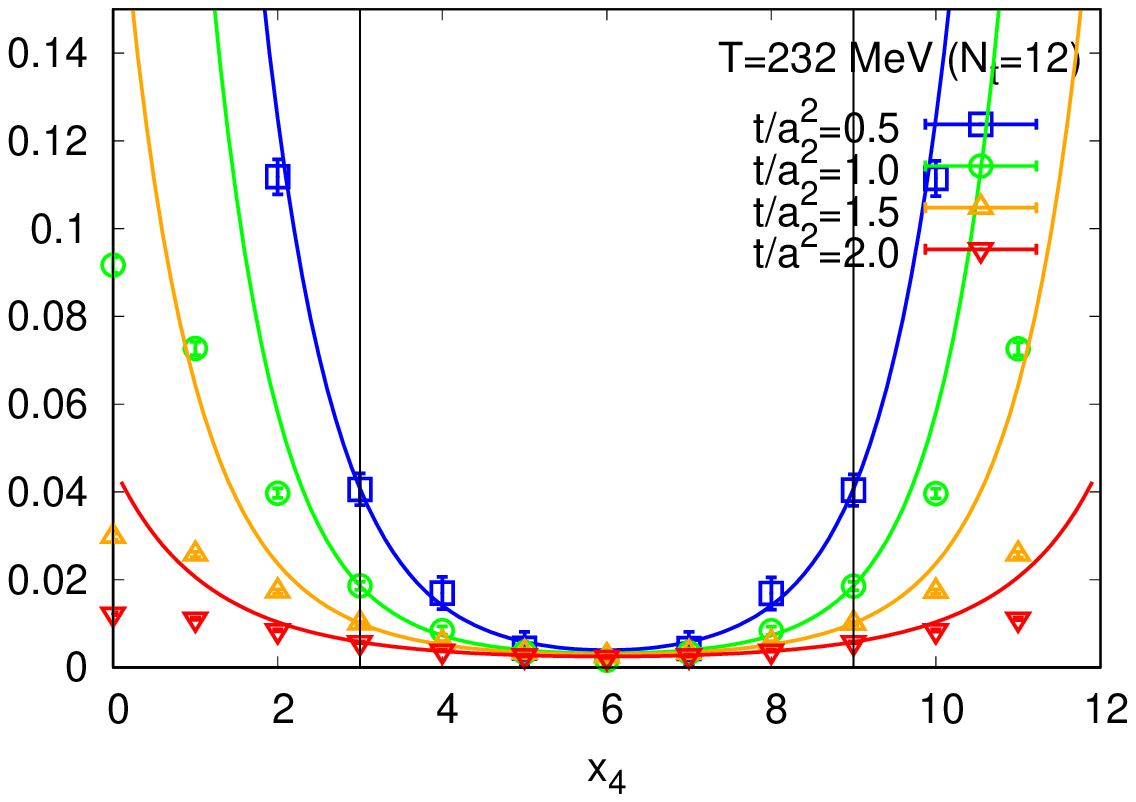}
  \includegraphics[width=4.9cm,trim=-20mm -20mm 0mm 0mm,clip]{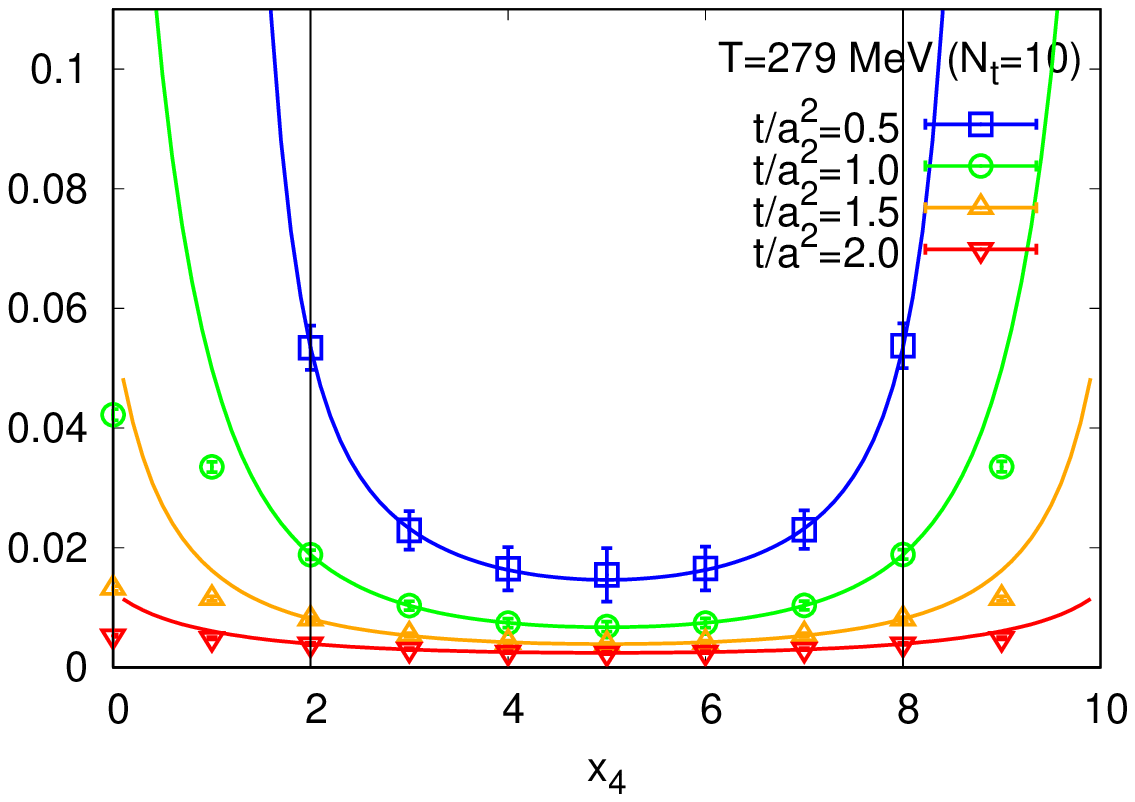}
  \includegraphics[width=4.9cm,trim=-20mm -20mm 0mm 0mm,clip]{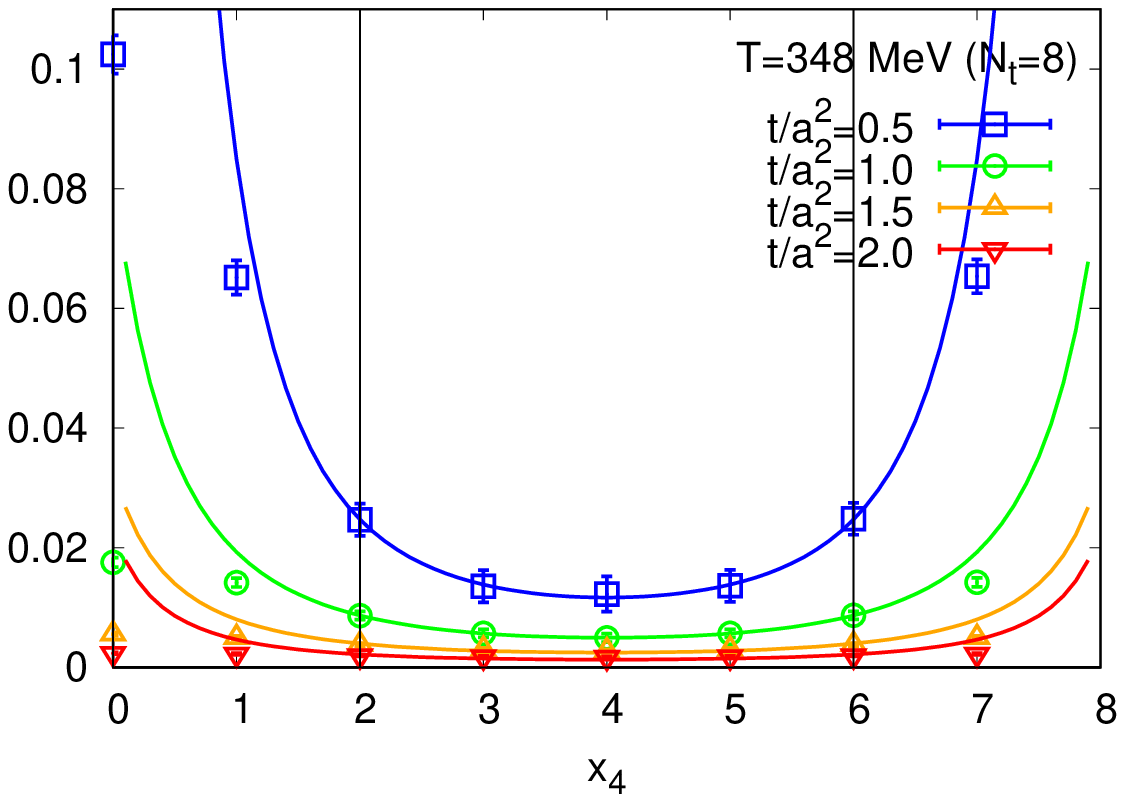}
  \includegraphics[width=4.9cm,trim=-20mm -20mm 0mm 0mm,clip]{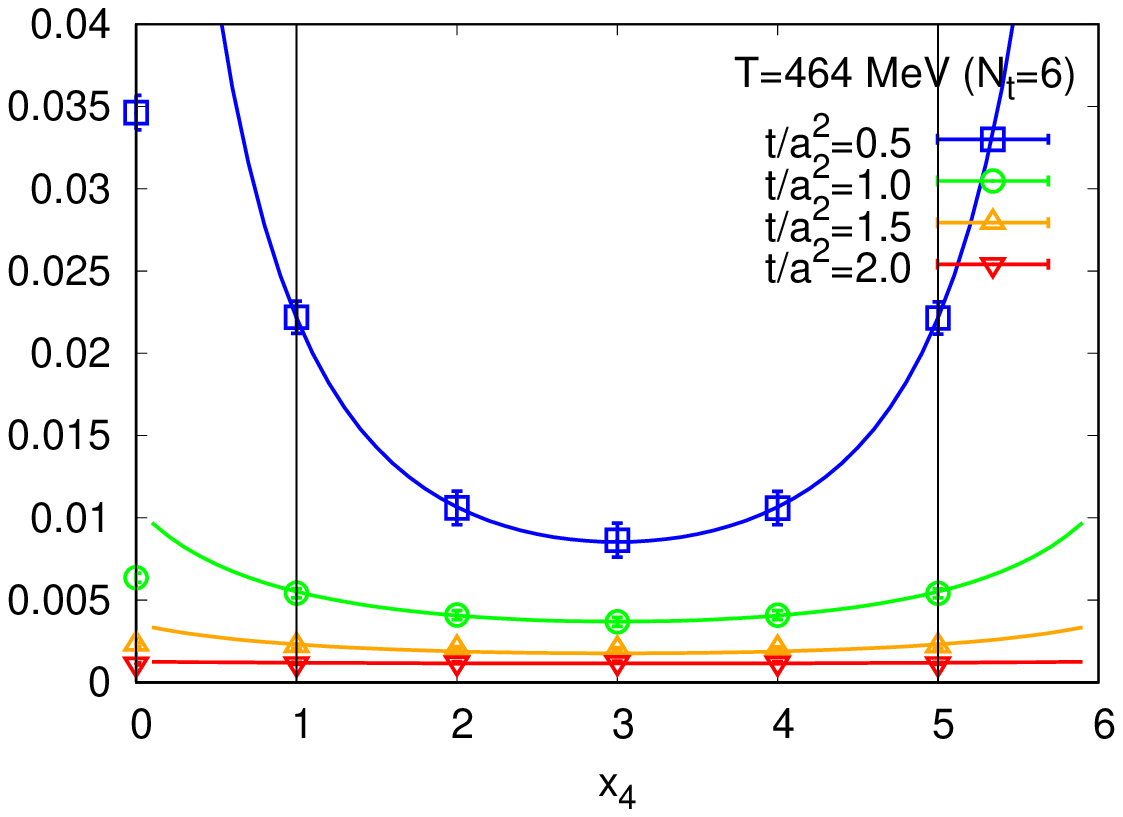}
  \vspace*{-3em}
  \caption{EMT correlation function $\sum_{\vec{x}}\vev{T_{ij}(t;\vec{x},x_{4})\,T_{ij}(t;0)}$ as function of Euclidean time $x_{4}$ at flow time $t/a^{2}=0.5$ (blue square), $1.0$ (green circle), $1.5$ (orange up triangle) and $2.0$ (red down triangle).
  From the top left to the bottom the temperature is $T\simeq174$, $199$, $232$, $279$, $348$ and $464$ MeV ($N_t=16$, $14$, $12$, $10$, $8$ and $6$, respectively).
  Solid lines are the fits with the Breit-Wigner fit ansatz using data between two dotted vertical lines.
}
\label{TTijij-x4-BW}
\end{figure}

\begin{figure}[tb]
 \centering
  \includegraphics[width=4.9cm,trim=-20mm -10mm 5mm 0mm,clip]{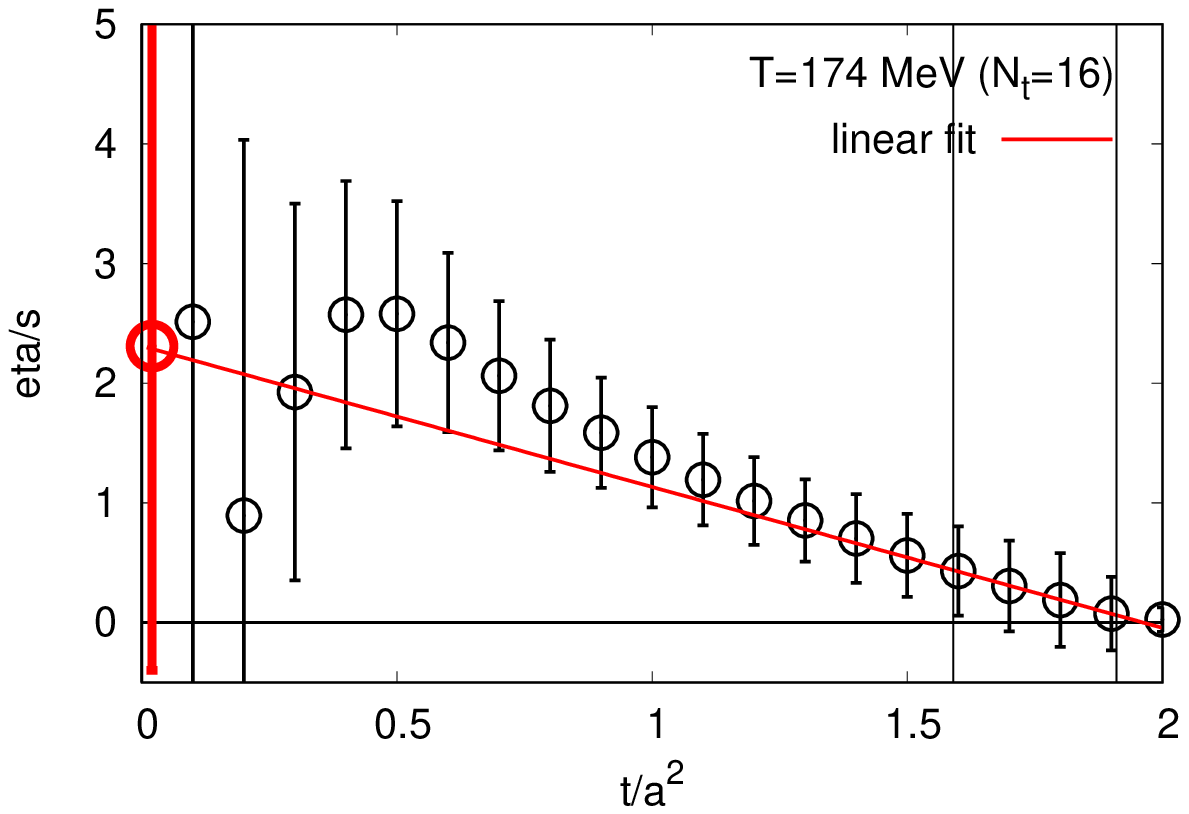}
  \includegraphics[width=4.9cm,trim=-20mm -10mm 5mm 0mm,clip]{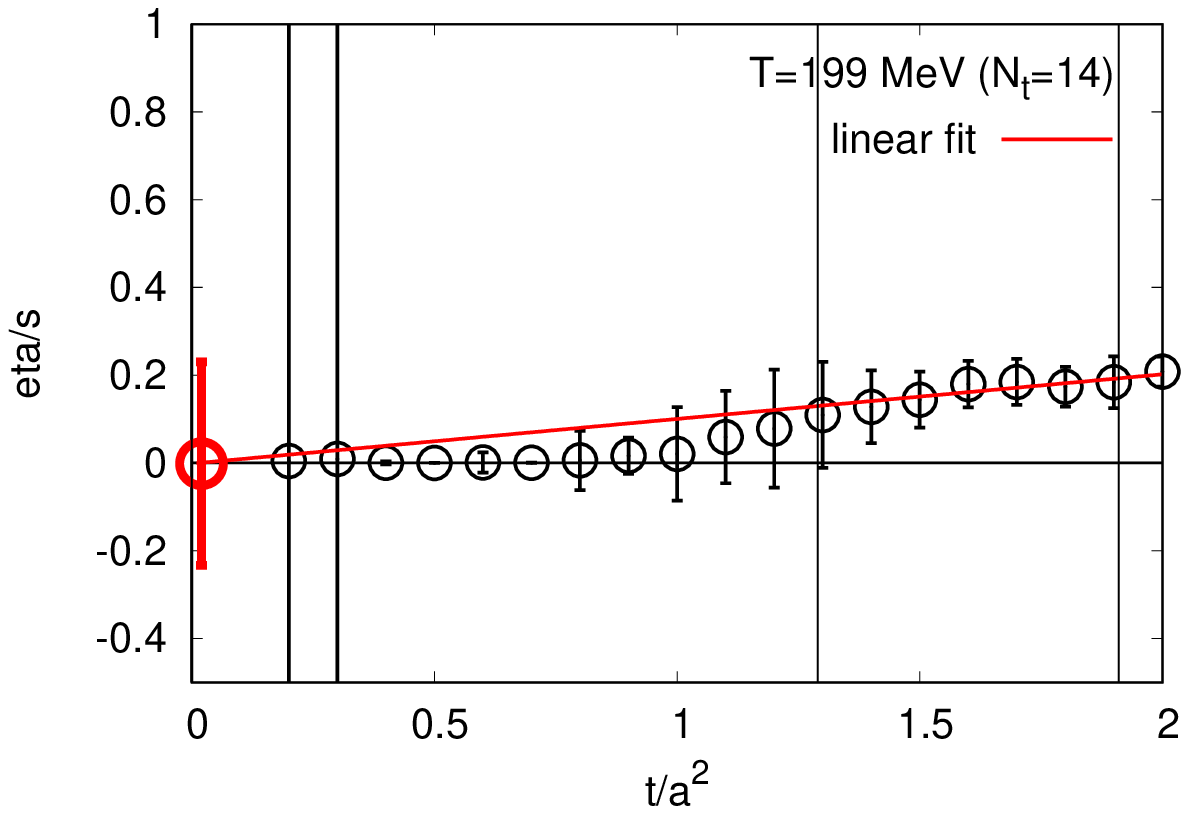}
  \includegraphics[width=4.8cm,trim=-20mm -10mm 5mm 0mm,clip]{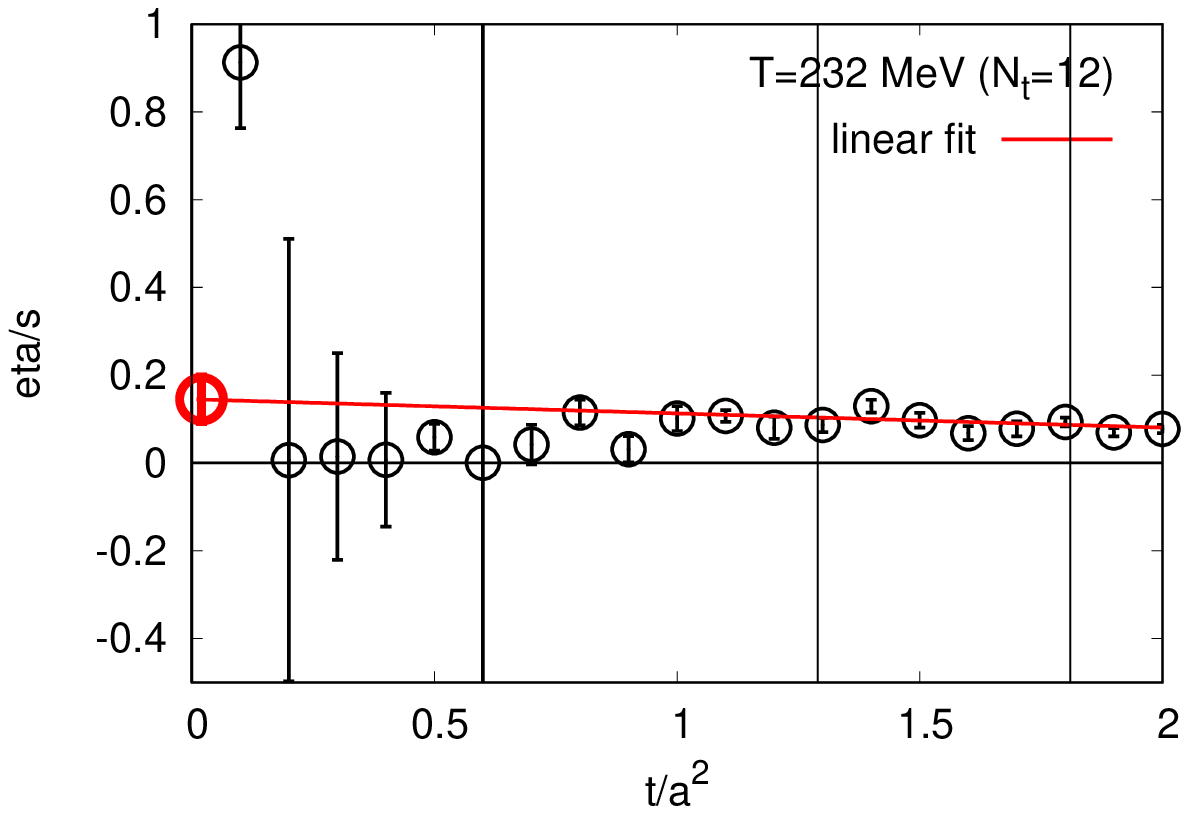}
  \includegraphics[width=4.9cm,trim=-20mm -20mm 5mm 0mm,clip]{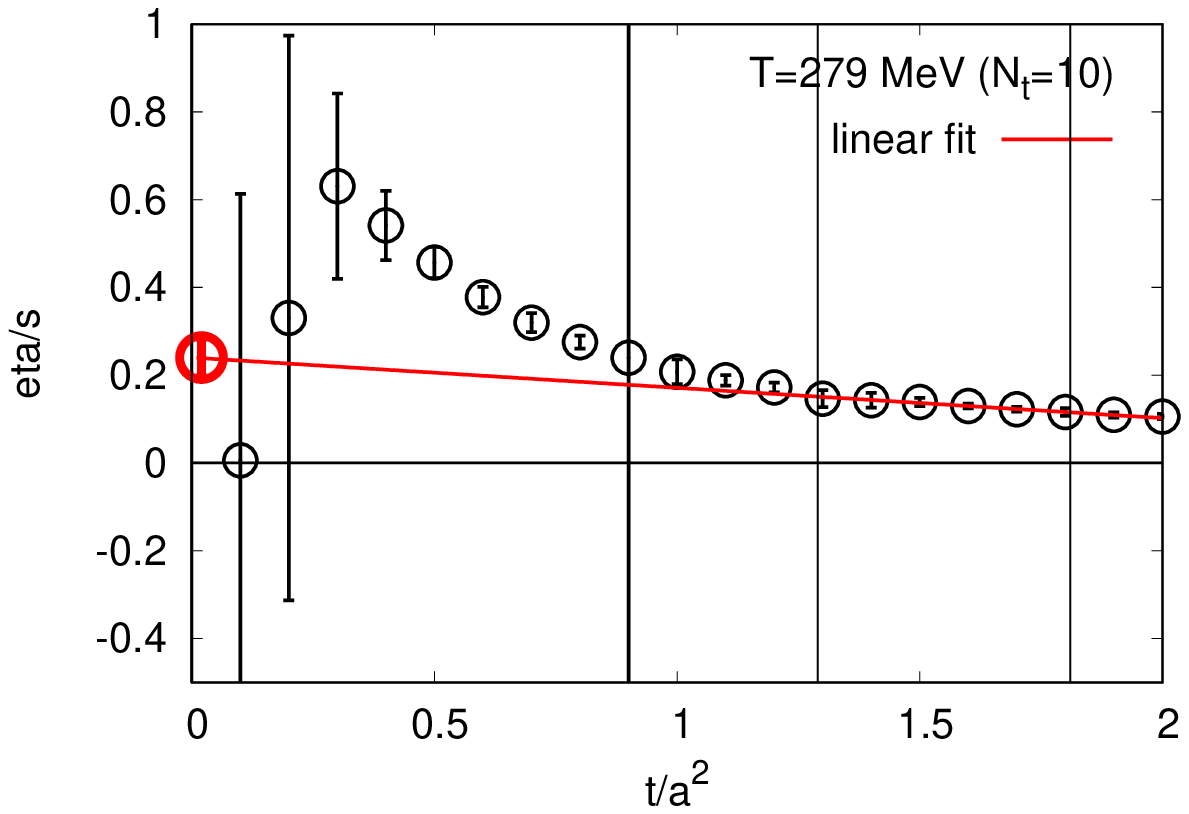}
  \includegraphics[width=4.9cm,trim=-20mm -20mm 5mm 0mm,clip]{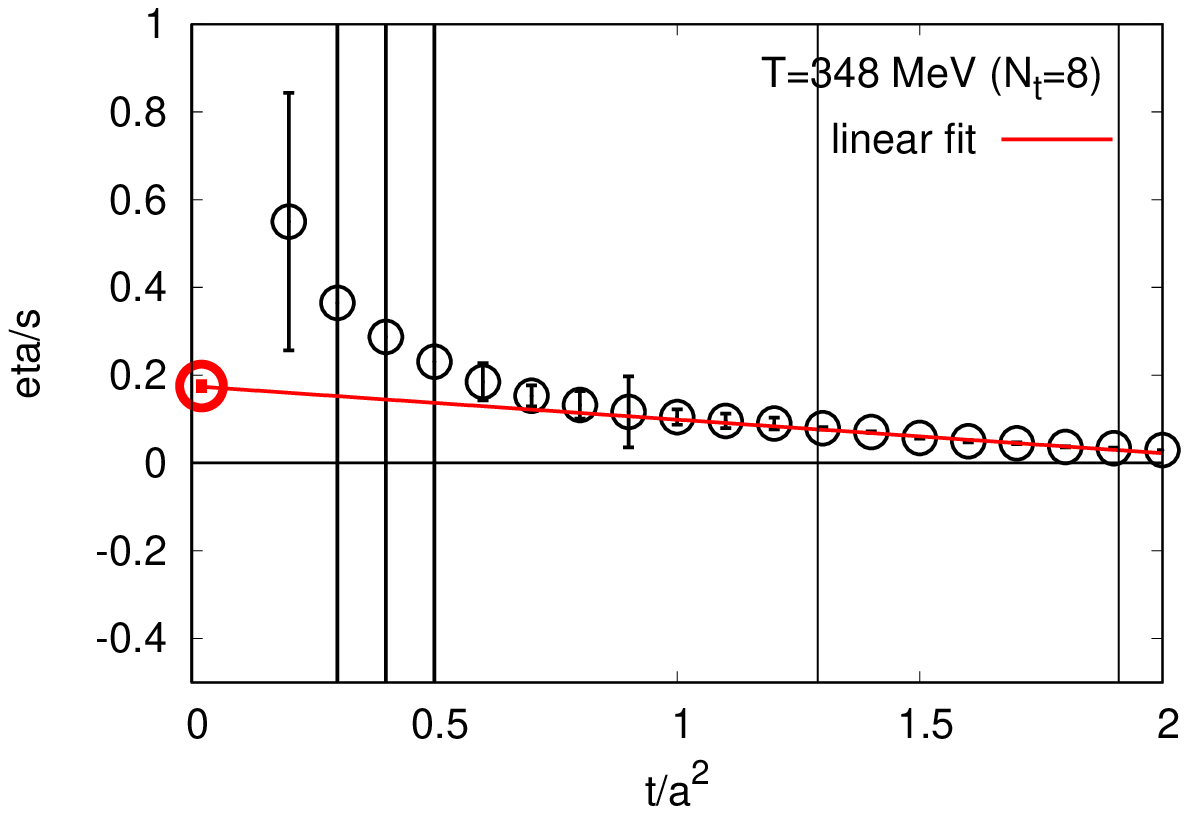}
  \includegraphics[width=4.9cm,trim=-20mm -20mm 5mm 0mm,clip]{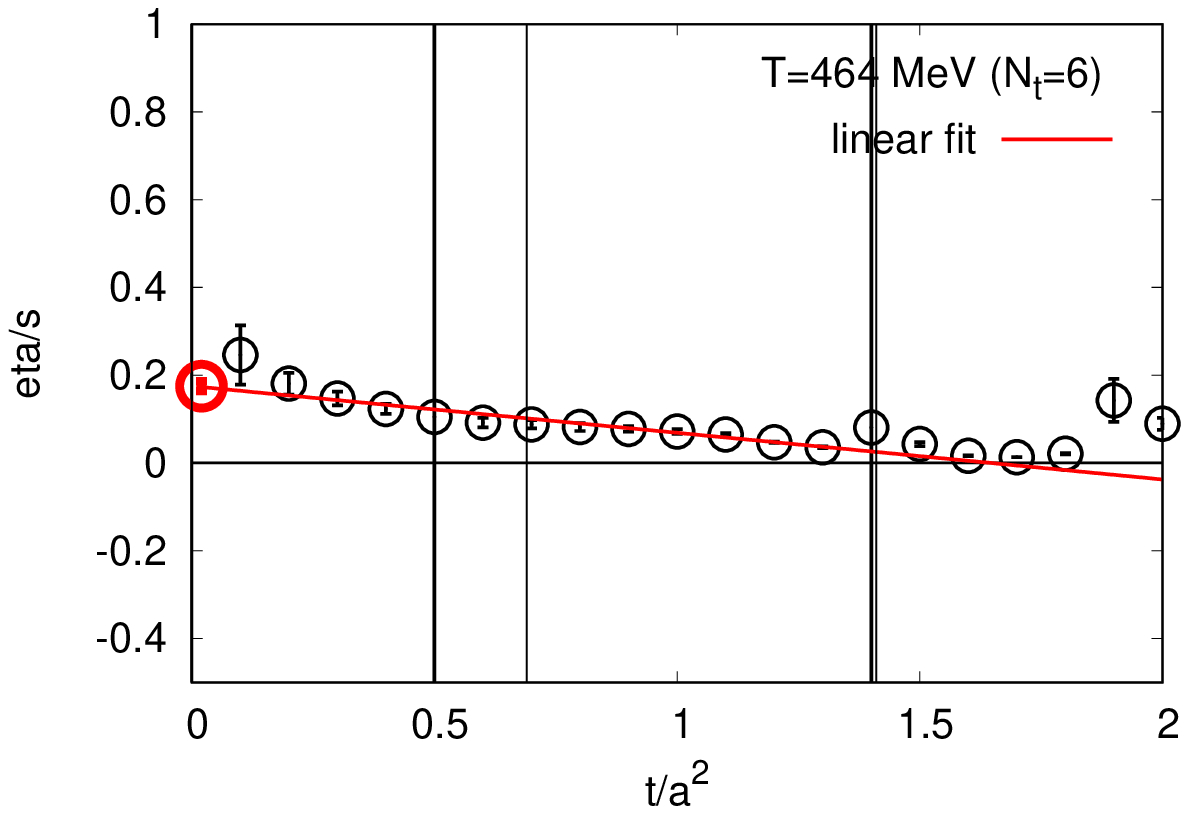}
  \vspace*{-3em}
  \caption{The shear viscosity divided by the entropy density $\eta/s$ as a function of the flow time.
  $\eta/s$ is extracted by fitting the correlation function with the Breit-Wigner ansatz.
  We take the $t\to0$ limit with a linear fit using the fit range indicated by two dotted vertical lines.
}
\label{etaovers1}
\end{figure}

\begin{figure}[tb]
 \centering
  \includegraphics[width=6.2cm,trim=-20mm -20mm 0mm 0mm,clip]{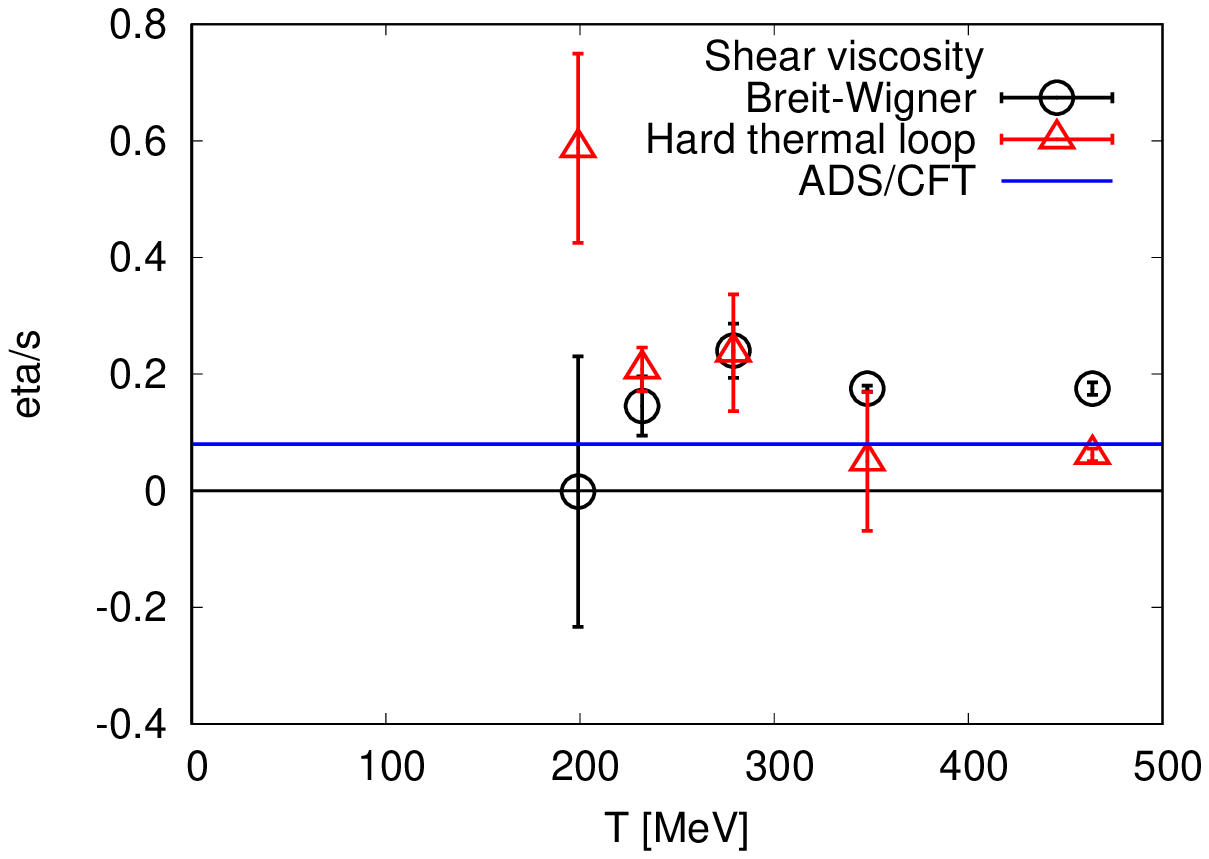}
  \hspace{5mm}
  \includegraphics[width=6.2cm,trim=-20mm -20mm 0mm 0mm,clip]{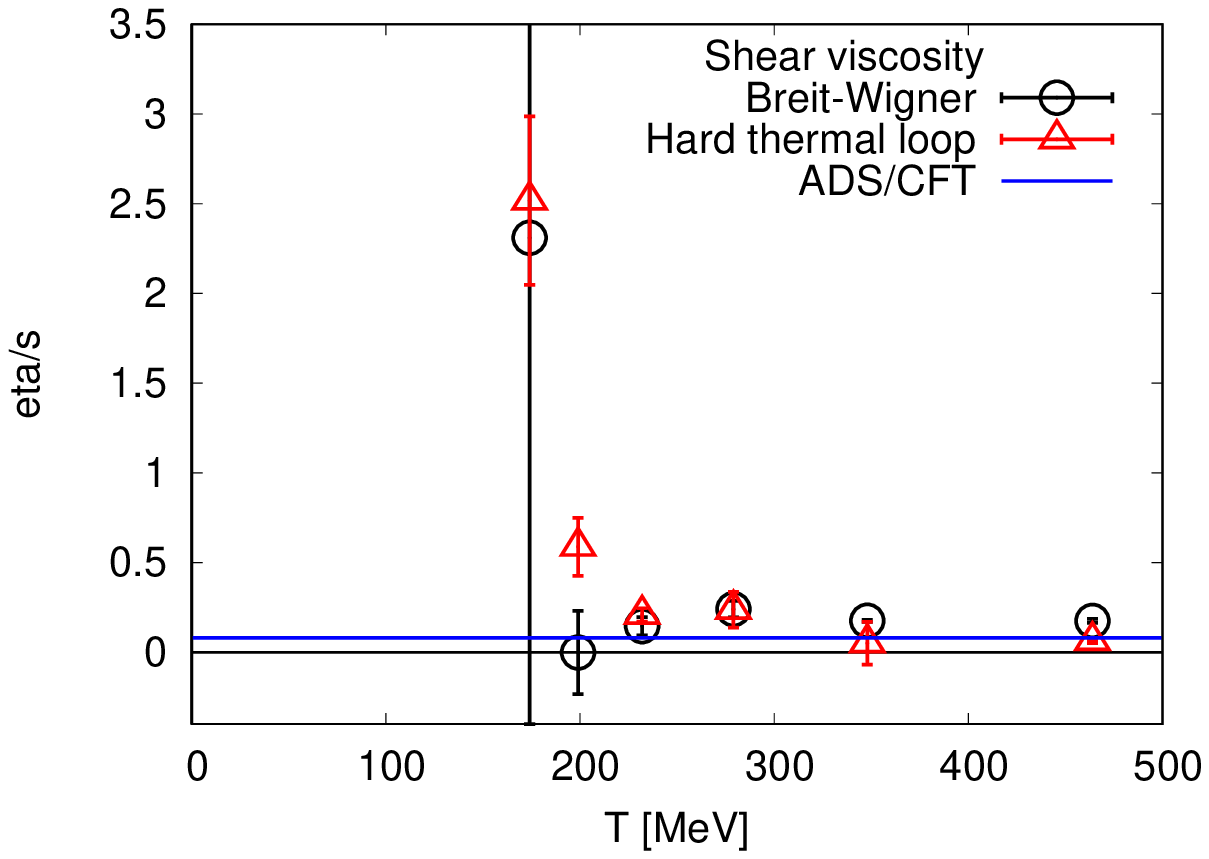}
  \vspace*{-3em}
  \caption{Preliminary results for the shear viscosity divided by the entropy density, $\eta/s$, as a function of the temperature.
  The viscosity is extracted by fitting the correlation function with the Breit-Wigner ansatz (black circles) and the hard thermal loop ansatz (red triangles).
  The blue solid line is $1/4\pi$ from the AdS/CFT correspondence.
  The right panel is the same but differs in vertical scale.
}
\label{etaovers-T}
\end{figure}

\begin{table}[b]
\caption{Preliminary results for the shear viscosity divided by the entropy density, $\eta/s$, at each temperature given by Breit-Wigner (BW) fit ansatz and hard thermal loop (HTL) fit ansatz.}
\label{table1}
 \centering
  \vspace*{0.5em} 
\begin{tabular}{|c|c|c|c|c|c|c|}
\hline
T[MeV] & $174$ & $199$ & $232$ & $279$ & $348$ & $464$ \cr
\hline
BW ansatz  
& $2.3(2.7)$ & $0.00(23)$ & $0.145(51)$ & $0.240(47)$ & $0.1753(53)$ & $0.175(11)$ \cr
HTL ansatz 
& $2.52(47)$ & $0.59(16)$ & $0.208(38)$ & $0.24(10)$ & $0.05(12)$ & $0.061(11)$ \cr
\hline
\end{tabular}
\label{default}
\end{table}%

In Fig.\ref{TTijij-x4-BW} we plot the shear correlation function  $\sum_{\vec{x}}\vev{T_{ij}(t;\vec{x},x_{4})T_{ij}(t;0)}$ as function of Euclidean time $x_{4}$ at four different flow times $t$ for six different temperatures $T$ shown in each panel.
Spatial indices are averaged over all $i\neq j$ combinations.
Within the fit range indicated by the vertical dotted lines, the Breit-Wigner ansatz \eqn{BW-model} fits the data well with $\chi^{2}/{\rm dof}<2$ at low temperatures $T\le279$ MeV.
On the other hand, $\chi^{2}/{\rm dof}$ exceeds ten at higher temperatures.

The results for the shear viscosity divided by the entropy density, $\eta/s$, is plotted in Fig.~\ref{etaovers1} as a function of the flow time at six temperatures.
We take the $t\to0$ limit by using data within the region shown by two vertical dotted lines.
Our preliminary results for $\eta/s$ by the Breit-Wigner ansatz at each temperature are summarized in the first row of Table~\ref{table1}, and are plotted in Fig.~\ref{etaovers-T} by black circles.

The same procedure works well also with the hard thermal loop fit ansatz \eqn{HTL-model}.
Our preliminary results for $\eta/s$ by the hard thermal loop ansatz are summarized in the second row of Table~\ref{table1}, and are plotted in Fig.~\ref{etaovers-T} by red triangles.

For the bulk viscosity $\zeta$, we find that our diagonal correlation function
$\sum_{\vec{x}}\vev{T_{ii}(t;\vec{x},x_{4})\,T_{ii}(t;0)}$ is statistically not fine enough to extract a non-trivial value of $\zeta/s$.
Both with Breit-Wigner and hard thermal loop fit ans\"atze, we find large $\chi^{2}/{\rm dof}$ above five.
The resultant values of $\zeta/s$ are consistent with zero with large statistical error for all temperatures we studied.

\section{Conclusion}

We calculated two point correlation functions of the energy-momentum tensor (EMT) in lattice QCD with $(2+1)$-flavors at finite temperature, $174\le T\le464$ MeV.
Nonperturbatively renormalized EMT is calculated by applying the gradient flow renormalization scheme.
By virtue of the gradient flow, statistical error is suppressed at finite flow time and the correlation functions are evaluated with a good precision.
We apply two types of  model fit for the correlation function to extract the spectral function.
The procedure works well for the shear correlation function $\sum_{\vec{x}}\vev{T_{ij}(t;\vec{x},x_{4})\,T_{ij}(t;0)}$.
Our preliminary results for the shear viscosity are given in Fig.~\ref{etaovers-T} and Table~\ref{table1} as a function of temperature.
The resultant shear viscosity seems to be consistent with the small experimental value given in Ref.~\cite{Gale:2013da} well above the pseudo-critical temperature $T_{\mathrm{pc}}\approx190\,\mathrm{MeV}$.
On the other hand, we need a higher statistics for the bulk viscosity.

\vspace{2mm}
This work was in part supported by JSPS KAKENHI Grant Numbers JP18K03607, JP17K05442, JP16H03982,  JP15K05041, 
JP26400251, JP26400244, and JP26287040.
This research used computational resources of COMA and Oakforest-PACS provided by the Interdisciplinary Computational Science Program of Center for Computational Sciences, University of Tsukuba,
Oakforest-PACS at JCAHPC through the HPCI System Research Project (Project ID:hp17208), OCTOPUS at Cybermedia Center, Osaka University, and ITO at R.I.I.T., Kyushu University.
The simulations were in part based on the Lattice QCD code set Bridge++ \cite{bridge}.

\end{document}